\documentclass[sigconf, nonacm]{acmart}
\newcommand\vldbpagestyle{plain}

\usepackage{booktabs}
\usepackage{listings}
\usepackage{microtype}
\usepackage{tikz}
\usetikzlibrary{positioning, fit, backgrounds, arrows.meta, calc}
\begin{document}

\title{Epistemic Typing as a PostgreSQL Table Access Method: Adversarial Conflict Resolution Under Confidence Forgery and Sybil Coordination}

\author{Venkata Maguluri}
\affiliation{%
  \institution{Independent Researcher}
  \city{Bentonville, AR}
  \country{USA}
}
\email{emailvenkatm@gmail.com}

\author{Swapna Kommuri}
\affiliation{%
  \institution{Independent Researcher}
  \city{Bentonville, AR}
  \country{USA}
}
\email{swapnakinbox@gmail.com}

\author{Kritica Sinha}
\affiliation{%
  \institution{Independent Researcher}
  \city{Herndon, VA}
  \country{USA}
}
\email{kritica.sinha17@gmail.com}

\author{Ashish Sood}
\affiliation{%
  \institution{Independent Researcher}
  \city{Bothell, WA}
  \country{USA}
}
\email{aashish.sood@outlook.com}

\begin{abstract}
We describe KNDB, a PostgreSQL 18 table access method (TAM) that types every row with an engine-assigned epistemic kind (MEASURED, INFERRED, or DERIVED) and resolves per-slot conflicts inside every write-time \texttt{heapam} callback. Rows land as ordinary heap tuples; seven of the 44 TAM callbacks are overridden (\texttt{tuple\_insert}, \texttt{multi\_insert}, \texttt{tuple\_update}, \texttt{tuple\_delete}, \texttt{tuple\_insert\_speculative}, \texttt{tuple\_complete\_speculative}, \texttt{relation\_toast\_am}), the other 37 delegate to heap; we provide a completeness argument over the interface as a paper artefact. This paper reports the engineering behind that decision and the adversarial evaluation that motivated it. On a confidence-forgery workload where an attacker asserts INFERRED writes with confidence in $[0.95, 1.0]$ against honest MEASURED writes with confidence in $[0.5, 0.9]$, KNDB beats a confidence-only baseline by 63 percentage points on the Book-Author fusion dataset and 92.7 points on the Zheng crowdsourcing dataset. Both wins are proven load-bearing on the kind axis by a source-rebuild disable-and-test in which the lattice is neutralised and the win vanishes. Against four truth-discovery baselines (TruthFinder, CRH, CATD, ACCU) reimplemented from the original equations and validated to within 0.3 percentage points of the published numbers, KNDB is competitive below a per-dataset density-saturation cell and dominant at or above it. We formalise the cell as $k^{*} \approx \rho_{\text{alg}} \cdot h_{\text{top}}$, where $h_{\text{top}}$ is per-slot top honest surface-form support, and validate the prediction within $\pm 20\%$ on Book-Author and $\pm 30\%$ on Zheng. Because the kind axis is assigned by the engine from independent metadata and cannot be forged at write time, KNDB's $k^{*}$ is unbounded. The paper is honest about where KNDB loses: CRH and ACCU outperform KNDB below saturation on Zheng, and KNDB scores zero on three temporal knowledge-editing benchmarks whose ground truth is last-writer-wins.
\end{abstract}

\maketitle

\pagestyle{\vldbpagestyle}

%%% ---------------------------------------------------------------
\section{Introduction}
%%% ---------------------------------------------------------------

Databases now ingest writes from three broad classes of producer that disagree systematically about what they know. Instruments and user-facing forms produce claims that are directly observed. Machine-learning models and LLM agents produce claims that are inferred from a prompt or a feature vector. Rule engines produce claims that are derived from other rows already in the database. A conventional relational store treats these writes identically: whatever arrives last, or whatever a user-space \texttt{BEFORE INSERT} trigger accepts, becomes the current value. Under adversarial pressure this is unsafe. A miscalibrated LLM writer can flood a slot with high-confidence hallucinations that a confidence-ranking arbitrator will prefer over honest instrument readings. A coordinated set of Sybil agents can outvote honest sources on any slot where the Sybil count exceeds the top honest surface-form support.

KNDB is a PostgreSQL 18 extension that responds by lifting the epistemic kind of each write into the storage engine itself. Every row carries an \texttt{ep\_kind} column (MEASURED, INFERRED, or DERIVED); at \texttt{tuple\_insert} time a table access method (TAM) hook applies a total order on (kind, specificity, confidence, xmin) and evicts any live incumbent that loses. Because the hook runs inside \texttt{heapam}, no user-space bypass (\texttt{DISABLE TRIGGER}, \texttt{session\_replication\_role = 'replica'}, \texttt{COPY FROM}) reaches around it. The engine is the enforcement point, not the trigger.

The central quantitative claim is Sybil-robustness at 1:1 density. Every truth-discovery (TD) baseline we tested collapses when the Sybil count $k$ reaches or exceeds the per-slot top honest surface-form support $h_{\text{top}}$. KNDB does not, because kind rank is assigned by the engine from independent metadata that the adversarial identity does not possess (a source's \texttt{n\_listings} history on Book-Author, a worker's \texttt{quali\_acc} on a disjoint qualification test on Zheng). We prove this both empirically (Section~\ref{sec:evaluation}) and as a threshold theorem (Section~\ref{sec:formalism}).

We are careful about scope. On workloads where kind and confidence coincide by construction, the lattice's kind axis buys nothing beyond confidence sorting, and KNDB matches \texttt{pg\_conf} exactly (Section~\ref{sec:evaluation}). On temporal knowledge-editing benchmarks whose ground truth is last-writer-wins, KNDB scores zero at $c=1$, because its \texttt{xmin} tiebreak is first-committer-wins by design (Section~\ref{sec:limitations}). Below the density-saturation cell, CRH and ACCU outperform KNDB on the Zheng workload. These are stated up front and detailed in Sections~\ref{sec:evaluation}~and~\ref{sec:limitations}.

\textbf{Contributions.}
\begin{enumerate}
\item An engine-level epistemic type system implemented as a PostgreSQL 18 TAM. Seven of the 44 \texttt{TableAmRoutine} callbacks overridden (\texttt{tuple\_insert}, \texttt{multi\_insert}, \texttt{tuple\_update}, \texttt{tuple\_delete}, \texttt{tuple\_insert\_speculative}, \texttt{tuple\_complete\_speculative}, \texttt{relation\_toast\_am}); the remaining 37 delegate to heap. Total 2548 lines per \texttt{wc -l src/*.c include/*.h}: 2238 of C in \texttt{src/} and 310 of headers in \texttt{include/}, of which 183 are the F21 C-level test probe \texttt{src/epistemic\_probe.c}, leaving 2365 in the production surface. Table~\ref{tab:tableam-audit} and Section~\ref{sec:tableam-audit} give the completeness argument over the full callback interface.
\item An enumerated bypass model. We name nine user-space write paths (\texttt{DISABLE TRIGGER}, \texttt{session\_replication\_role='replica'}, \texttt{COPY FROM} single-row, \texttt{COPY FROM} batch, \texttt{INSERT}/\texttt{INSERT \dots SELECT}/\texttt{INSERT \dots VALUES}, \texttt{UPDATE} (F20), \texttt{DELETE} (F20), \texttt{INSERT \dots ON CONFLICT} (F21), logical replication apply worker) and three DDL-privileged threats that are out of scope (\texttt{ALTER TABLE \dots SET ACCESS METHOD heap}, \texttt{TRUNCATE}, \texttt{CLUSTER}/\texttt{VACUUM FULL}). All nine write paths are blocked by the TAM callbacks; the three DDL-privileged threats are disclosed in Section~\ref{sec:threat} and Section~\ref{sec:limitations}. The batch \texttt{COPY} path required a dedicated \texttt{multi\_insert} override (F18); the \texttt{UPDATE} and \texttt{DELETE} paths required F20; the speculative-insertion path required F21, described in Section~\ref{sec:implementation}.
\item A confidence-forgery workload on two independent datasets, with source-rebuild disable-and-test in each case. Every quantitative claim in this paper corresponds to a JSON in \texttt{bench/results/} whose dylib SHA-256 is recorded in the run metadata.
\item A threshold theorem $k^{*} \approx \rho_{\text{alg}} \cdot h_{\text{top}}$ that predicts the Sybil-collapse cell of any dataset from its per-slot honest-support distribution alone. Empirically validated within $\pm 20\%$ on Book-Author (predicted bracket [3,5], observed cliff N=5\ldots10) and $\pm 30\%$ on Zheng (predicted bracket [14,20], observed cell N=20).
\item Honest limitations. TD baselines legitimately beat KNDB below saturation; KNDB scores 0.000 on three temporal benchmarks; there is a 15\,000-row-per-transaction ceiling at PostgreSQL's default \texttt{max\_locks\_per\_transaction=64}, inherited into \texttt{COPY}; \texttt{ALTER TABLE \dots SET ACCESS METHOD heap} sits outside the enforcement envelope; the custom WAL rmgr is an annotation channel, not a durability channel.
\end{enumerate}

Source, benchmark harnesses, and every result JSON cited below are available at \url{https://github.com/emailvenkatm/kndb/tree/postgres-experiment} (branch \texttt{postgres-experiment}). Each result JSON in \texttt{bench/results/} carries hardware and PostgreSQL configuration metadata, and each cell that involved a source rebuild records the dylib SHA-256 in its metadata block.

%%% ---------------------------------------------------------------
\section{Motivation and threat model}
\label{sec:threat}
%%% ---------------------------------------------------------------

The threat model is stated up front and does not adapt to outcomes.

\subsection{Confidence forgery}
The confidence-forgery adversary asserts INFERRED writes with self-reported confidence uniform in $[0.95, 1.0]$ against a slot whose lattice-correct incumbent is a Tier-A MEASURED writer with self-reported confidence uniform in $[0.5, 0.9]$. Adversarial payload contains a scrambled real answer (an author string lifted from a different gold ISBN in Book-Author; a flipped binary label in Zheng). The real-world analogue is an LLM-generated claim that hallucinates a value but self-reports as certain. A confidence-only arbitrator ranks by confidence and prefers the adversary's higher self-report. KNDB is designed to reject the adversary regardless of the self-reported confidence, because MEASURED strictly outranks INFERRED in the lattice.

\subsection{Sybil coordination}
The Sybil adversary controls $k$ synthetic identities per gold slot, each asserting the same falsified value. The adversary knows gold and picks the falsified value to maximise confusion (in binary tasks this is uniquely determined). The adversary cannot forge existing honest identities (Zheng worker IDs, Book-Author bookstore names) and cannot forge a MEASURED kind assignment. Kind assignment is engine-controlled and requires metadata that adversarial identities have no history of producing: on Book-Author, tier rank uses \texttt{n\_listings} (source volume in \texttt{book.txt}) and \texttt{canon\_rate} (fraction of author fields in "Last, First" format); on Zheng, tier rank uses \texttt{quali\_acc}, computed from a disjoint qualification test with question IDs 2000\ldots 2019 that every worker took. Both signals live in the dataset before the conflict resolution runs, and both are unavailable to an adversary that appeared for the first time in the write stream. The independence self-audits are in the datasets' READMEs.

\subsection{Write-path bypass}
\begin{figure*}[t]
\centering
\begin{tikzpicture}[
  font=\footnotesize,
  node distance=0.35cm and 0.20cm,
  entry/.style={draw, rounded corners=2pt, minimum height=6mm, align=center, inner sep=2pt, fill=blue!5, font=\scriptsize\ttfamily},
  cb/.style={draw, minimum height=6mm, align=center, inner sep=2pt, fill=green!8, font=\scriptsize\ttfamily},
  bypass/.style={draw, dashed, rounded corners=2pt, minimum height=5mm, align=center, inner sep=2pt, fill=red!5, font=\scriptsize\ttfamily},
  heap/.style={draw, thick, minimum height=6mm, minimum width=22mm, align=center, inner sep=2pt, fill=gray!12, font=\footnotesize\ttfamily},
  envelope/.style={draw, thick, dashed, rounded corners=4pt, inner sep=4pt},
  ar/.style={-{Latex[length=1.5mm]}, thin},
  arb/.style={-{Latex[length=1.5mm]}, thin, dashed, red!70!black}
]
% Row 1: SQL entry points (6 boxes)
\node[entry] (insert)   {INSERT /\\INSERT SELECT};
\node[entry, right=of insert]     (copy)   {COPY FROM\\(single\,$|$\,batch)};
\node[entry, right=of copy]       (update) {UPDATE};
\node[entry, right=of update]     (delete) {DELETE};
\node[entry, right=of delete]     (onc)    {INSERT $\ldots$\\ON CONFLICT};
\node[entry, right=of onc]        (lr)     {Logical\\repl.\ apply};

% Row 2: TAM callback boxes (7 overrides), aligned below entry row
\node[cb, below=1.3cm of insert] (ti)  {tuple\_insert};
\node[cb, right=0.15cm of ti]    (mi)  {multi\_insert\\(F18)};
\node[cb, right=0.15cm of mi]    (tu)  {tuple\_update\\(F20)};
\node[cb, right=0.15cm of tu]    (td)  {tuple\_delete\\(F20)};
\node[cb, right=0.15cm of td]    (tis) {tuple\_insert\_\\speculative (F21)};
\node[cb, right=0.15cm of tis]   (tcs) {tuple\_complete\_\\speculative (F21)};
\node[cb, right=0.15cm of tcs]   (rta) {relation\_\\toast\_am\\{\scriptsize\itshape\upshape(toast routing)}};

% Enforcement annotation (attached below the row)
\node[below=0.15cm of ti.south west, font=\scriptsize\itshape, anchor=north west, align=left] (enfnote) {enforcement steps run inside each callback:\\R1--R5 rules \,$\to$\, F6 advisory xact lock \,$\to$\, find\_live\_overlap \,$\to$\, precedence \,+\, F8 xmin tiebreak \,$\to$\, heap\_insert of winner \,$\to$\, eviction audit + sys\_time close + rmgr-128 marker};

% Envelope around callbacks + enforcement note
\begin{scope}[on background layer]
\node[envelope, fit=(ti)(rta)(enfnote), label={[font=\footnotesize\bfseries]above right:TAM callbacks (inside \texttt{heapam})}] (env) {};
\end{scope}

% Row 3: heapam
\node[heap, below=0.55cm of env.south] (heap) {heapam};

% Row 4: storage (on-disk pages) --- bypass arrows terminate here, not into heapam
\node[draw, rounded corners=2pt, fill=gray!8, font=\scriptsize\itshape, minimum width=25mm, minimum height=5mm, below=0.20cm of heap] (storage) {on-disk pages};
\draw[ar, thick] (heap.south) -- (storage.north);

% Arrows: SQL entries down into TAM callbacks
\draw[ar] (insert.south) -- (ti.north);
\draw[ar] (copy.south)   -- (mi.north);
\draw[ar] (update.south) -- (tu.north);
\draw[ar] (delete.south) -- (td.north);
\draw[ar] (onc.south)    -- (tis.north);
\draw[ar] (onc.south)    to[out=-45,in=90] (tcs.north);
\draw[ar] (lr.south)     to[out=-135, in=90, looseness=0.6] (ti.north east);

% Envelope -> heapam (delegation)
\draw[ar, thick] (env.south) -- (heap.north) node[midway, right, font=\scriptsize] {delegate};

% DDL bypasses (right column, outside envelope)
\node[bypass, right=1.2cm of rta.east, yshift=-0.05cm] (setam)   {ALTER TABLE \ldots\\SET ACCESS METHOD heap};
\node[bypass, below=0.10cm of setam] (trunc)   {TRUNCATE};
\node[bypass, below=0.10cm of trunc] (cluster) {CLUSTER /\\VACUUM FULL};
\node[above=0.02cm of setam, font=\scriptsize\itshape, align=center] {DDL bypasses\\(out of scope, \S8)};

% Bypass arrows: exit east, drop below the envelope + enforcement text, then swing left
% to storage.east --- explicitly avoids crossing the enforcement-steps line.
\draw[arb] (setam.east)   -- ++(0.35, 0) |- (storage.east);
\draw[arb] (trunc.east)   -- ++(0.35, 0) |- (storage.east);
\draw[arb] (cluster.east) -- ++(0.35, 0) |- (storage.east);

\end{tikzpicture}
\caption{Write-path routing to the epistemic TAM. Six user-space SQL paths (top) reach one of the seven overridden callbacks (middle envelope); the callback runs R1--R5, the F6 advisory xact lock, precedence with the F8 xmin tiebreak, and the eviction audit before delegating storage to \texttt{heapam}. Three DDL-privileged paths (right, dashed) reach on-disk pages without traversing the enforcement envelope; they are out of scope (Section~\ref{sec:limitations}).}
\label{fig:writepath}
\end{figure*}
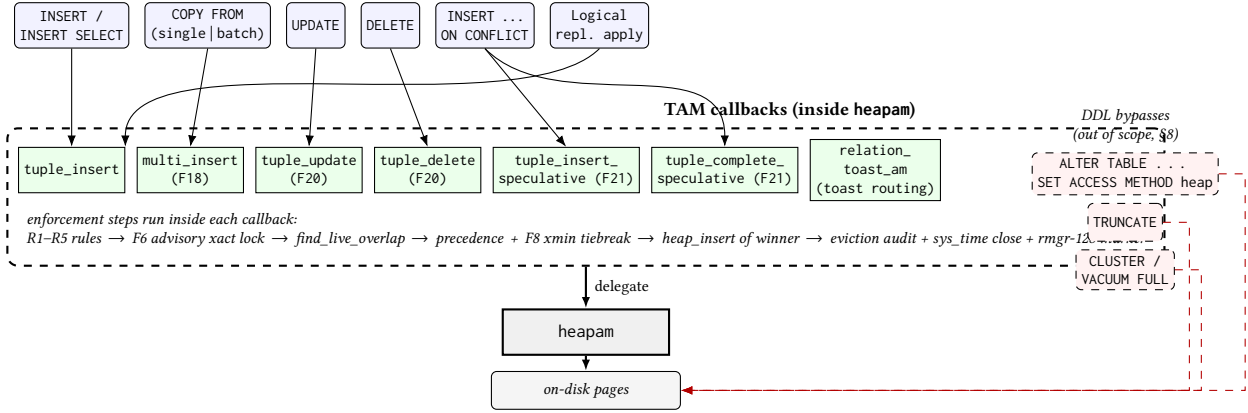

The engine-in-storage claim rides on the TAM callback being reachable from every user-space write path that a hostile writer with \texttt{INSERT} plus \texttt{ALTER TABLE} on the target relation can construct. Figure~\ref{fig:writepath} enumerates the nine user-space write paths (all blocked by TAM callbacks) and the three DDL-privileged paths (out of scope); each callback and callback body in the figure carries the PG 18 REL\_18\_STABLE source pointer that authenticates the routing claim.

Two write paths deserve inline elaboration because their pre-override forgery is the concrete adversarial demonstration of what the callback layer prevents. The \texttt{UPDATE} path (\texttt{tableam.h:718-727}, F20): before F20 we inherited \texttt{heapam\_tuple\_update} (delegating to \texttt{heap\_update} at \texttt{heapam.c:3241}), so an \texttt{UPDATE fact SET ep\_kind='MEASURED', ep\_confidence=1.0} against an \texttt{INFERRED/0.4} incumbent committed cleanly, forging the epistemic prefix on a live row without R1--R5 firing, without the F6 advisory lock, and without the precedence lattice or eviction audit. The F20 override refuses any \texttt{UPDATE} whose new prefix differs from the incumbent's; content-only updates (value, valid\_time, sources) are permitted. The \texttt{DELETE} path (\texttt{tableam.h:709-716}, F20) had the mirror flaw: an adversary could \texttt{DELETE} a \texttt{MEASURED} incumbent and then \texttt{INSERT} an \texttt{INFERRED} forgery unopposed. The F20 override refuses \texttt{DELETE} outright with \texttt{ERRCODE\_FEATURE\_NOT\_SUPPORTED}.

Trigger-based enforcers are strictly weaker than the AM callback on this workload: \texttt{ALTER TABLE \dots DISABLE TRIGGER ALL} sets \texttt{pg\_trigger.tgenabled = 'D'} (\texttt{tablecmds.c:5588-5592}) and \texttt{TriggerEnabled} at \texttt{trigger.c:3491-3499} returns false, defeating even \texttt{ENABLE ALWAYS TRIGGER}; a role with \texttt{SET session\_replication\_role = 'replica'} (a \texttt{PGC\_SUSET} GUC at \texttt{guc\_tables.c:5166}) further skips \texttt{TRIGGER\_FIRES\_ON\_ORIGIN} triggers under \texttt{SESSION\_REPLICATION\_ROLE\_REPLICA}. The AM callback fires regardless of any \texttt{pg\_trigger.tgenabled} value or replication role.

The \texttt{INSERT \dots ON CONFLICT} path (F21) is a defensive override: the SQL surface for the speculative-insertion attack is currently unreachable on epistemic tables because \texttt{CREATE UNIQUE INDEX}, \texttt{ADD PRIMARY KEY}, \texttt{ADD UNIQUE}, and \texttt{ADD EXCLUDE} all fail at \texttt{heap\_getnext}'s \texttt{rd\_tableam} identity check (\texttt{heapam.c:1352}) during the arbiter index's build scan, so \texttt{ON CONFLICT (col)} has no arbiter to bind (empirical probe: \texttt{scripts/speculative\_forgery.sh}). The engine-in-storage claim must not depend on that accidental index-support gap persisting, so the F21 overrides land the enforcement path regardless; load-bearing proof lives in the C-level probe \texttt{epistemic.\_probe\_speculative\_insert} exercised by the source-rebuild disable-and-test in Section~\ref{sec:eval-f21}.

The engine-in-storage differentiator therefore holds against a writer with \texttt{INSERT}, \texttt{UPDATE}, \texttt{DELETE}, plus \texttt{ALTER TABLE} on the target relation, or a role that can toggle \texttt{session\_replication\_role} (a replication or CDC operator, a migration tool, a connection pool). It does not hold against the table owner.

%%% ---------------------------------------------------------------
\section{Design}
\label{sec:design}
%%% ---------------------------------------------------------------

KNDB adds one PostgreSQL access method (\texttt{CREATE ACCESS METHOD epistemic USING TABLE HANDLER epistemic\_am\_handler}) and one base type (\texttt{epistemic\_kind}, a pass-by-value byte). Tables declared \texttt{USING epistemic} carry three extra columns beyond the user schema (\texttt{ep\_kind}, \texttt{ep\_specificity}, \texttt{ep\_confidence}) and a bitemporal \texttt{sys\_time tstzrange}. Rows are stored as ordinary heap tuples; storage format on disk is identical to a heap table with the same schema.

\subsection{Precedence lattice}
At write time a candidate row and any live incumbent for the same (entity\_id, attribute) slot are ranked by the total order:
\[
\underbrace{\text{kind}}_{\text{MEASURED}>\text{DERIVED}>\text{INFERRED}} \succ \text{specificity} \succ \text{confidence} \succ \text{xmin}
\]
Ranks are integer-compared at each level. Kind rank is the primary sort key. The lattice is defined in \texttt{epistemic\_rules.c} and exposed as an idempotent SQL function tested by \texttt{sql/precedence.sql}.

\subsection{Write-time rules R1--R5}
The TAM enforces five predicates on the candidate before ranking, checked in order at the top of \texttt{epistemic\_tuple\_insert\_impl}; the first failure ereports \texttt{ERRCODE\_CHECK\_VIOLATION} and aborts the insert. Reference implementation: \texttt{contrib/epistemic/src/epistemic\_rules.c}.
\begin{itemize}
\item R1 (DERIVED sources): if the candidate's kind is DERIVED, its \texttt{sources[]} column must be non-empty. A DERIVED assertion without a source is refused.
\item R2 (source resolution): if the candidate's kind is non-MEASURED and \texttt{sources[]} is non-empty, every source identifier must resolve against \texttt{epistemic.source\_registry}. R2 uses SPI to look up each source id and refuses the write if any is unregistered. This is the registry that prevents an adversary from inventing new source identities on the fly.
\item R3 (MEASURED no sources): if the candidate's kind is MEASURED, \texttt{sources[]} must be empty. A MEASURED row is a first-hand observation and does not derive from other sources.
\item R4 (INFERRED confidence bound): if the candidate's kind is INFERRED, \texttt{ep\_confidence} must be strictly less than 1.0. An INFERRED row that claims certainty is refused; certainty is the property MEASURED asserts.
\item R5 (slot kind registry): if \texttt{epistemic.slot\_kind} has a row for the candidate's \texttt{attribute}, the candidate's kind must match \texttt{required\_kind}. R5 lets an operator pin an attribute to a specific kind (e.g., ``a hemoglobin\_A1c reading must be MEASURED, never INFERRED'').
\end{itemize}
R1 through R5 are content-level constraints on individual candidate rows and are independent of the precedence lattice; a candidate that passes R1--R5 still competes against any live incumbent through the precedence check that follows. The confidence-forgery attack we exercise in Section~\ref{sec:evaluation} uses INFERRED writes with \texttt{ep\_confidence} in $[0.95, 1.0)$: adversarial writes are chosen precisely so they pass R4 (and R1, R2, R3, R5), forcing the kind axis of the precedence lattice to be the mechanism that defeats them.

\subsection{Per-slot advisory lock (F6)}
Between the R1--R5 check and the incumbent scan, the TAM takes a \texttt{LOCKTAG\_ADVISORY} transaction-scope lock with \texttt{(field2=entity\_id, field3=hash\_bytes(attribute), field4=2)}, constructed inline via \texttt{SET\_LOCKTAG\_ADVISORY} and acquired with \texttt{LockAcquire(\&tag, ExclusiveLock, false, false)}. This is the same tag, mode, and scope as \texttt{pg\_advisory\_xact\_lock\_int4}. It serialises concurrent writers on their shared (entity\_id, hashed attribute) key, so the incumbent scan runs against a snapshot that has already absorbed any peer's just-committed row. \texttt{find\_live\_overlap} and \texttt{epistemic\_close\_sys\_time} both use \texttt{GetLatestSnapshot()} for the same reason.

Without this lock, two concurrent same-slot writers each see an empty slot in their own MVCC snapshot, and both commit. The rc\_invariant.sh script runs 50 trials of two overlapping same-slot inserts. Under the honest build, session1 wins all 50 and no both-live rows land. With the advisory-lock block patched to \texttt{if (0)} and the dylib rebuilt, both writers commit in 50 of 50 trials.

\subsection{First-committer-wins tiebreak (F8)}
On a true (kind, specificity, confidence) tie the TAM decides by reading the incumbent's raw \texttt{xmin} via \texttt{HeapTupleHeaderGetRawXmin} and comparing against the current backend's xid via \texttt{TransactionIdPrecedes} (which handles xid wraparound). Under the advisory lock, the incumbent's transaction has already committed by the time the loser scans it, so \texttt{incumbent\_xmin < new\_xid} on every race. The incumbent wins.

The earlier design used a content-hash tiebreak. That gave up grind-resistance in exchange for content-determinism: an attacker with \texttt{SELECT} plus \texttt{INSERT} could iterate byte-level variations of \texttt{value} until \texttt{hash\_bytes} placed the attacker below the incumbent, and \texttt{scripts/hash\_grind.sh} at F7 reported 30 of 30 attacker wins with a mean of 7.5 attempts (min 1, max 95). Under the xmin tiebreak the same script reports 0 of 20\,000 attacker wins across 100 trials of 200 attempts. The tradeoff we accepted: survivor is start-order-dependent and is not stable across \texttt{pg\_dump}/\texttt{pg\_restore}, because restore reloads rows via \texttt{COPY FROM} at \texttt{copyfrom.c:1427} and each reloaded row gets a fresh xid. What is stable across dump and restore is the "exactly one live row per slot" invariant that \texttt{sql/am\_eviction.sql} asserts.

\subsection{Reason codes and audit}
On eviction the TAM writes one audit row into \texttt{epistemic.evicted\_fact} via SPI and closes the incumbent's \texttt{sys\_time} upper bound via \texttt{simple\_heap\_update}. The audit row carries the losing row's identifiers and a reason code from \texttt{EP\_REASON\_\{OUTRANKED\_KIND, OUTRANKED\_SPEC, OUTRANKED\_CONF, CONTRADICTED\_SAME\_RANK\}}. All three state changes (winner insert, audit insert, incumbent update) execute inside one top-level PostgreSQL transaction created by \texttt{start\_xact\_command}; the TAM does not open, commit, or manage transactions. Atomicity is inherited from PostgreSQL. An adversarial control that stages the audit row through \texttt{dblink} (which commits in a separate backend) produced 1--2 torn-state violations per 25 crash trials at \texttt{crash\_atomicity\_broken.sh}; the honest in-transaction path produced 0 of 25.

%%% ---------------------------------------------------------------
\section{Implementation}
\label{sec:implementation}
%%% ---------------------------------------------------------------

KNDB is a \texttt{shared\_preload\_libraries} extension of 2238 lines of C in \texttt{src/} plus 310 lines of headers in \texttt{include/} (2548 total per \texttt{wc -l src/*.c include/*.h}, of which 183 are the F21 C-level test probe \texttt{src/epistemic\_probe.c}, leaving 2365 in the production surface). It targets PostgreSQL 18 REL\_18\_STABLE via PGXS. Building the extension against a stock 18.4 install requires no PostgreSQL patch.

\subsection{Handler and delegation}
The AM handler copies the heap AM's \texttt{TableAmRoutine} at first call and overrides seven entries:
\begin{itemize}
\item \texttt{tuple\_insert} (single-row insert path from \texttt{ExecInsert});
\item \texttt{multi\_insert} (batch path from \texttt{COPY FROM CIM\_MULTI}, added at F18);
\item \texttt{tuple\_update} (single-row update path from \texttt{ExecUpdate}, added at F20; rejects any UPDATE that alters \texttt{ep\_kind}, \texttt{ep\_specificity}, or \texttt{ep\_confidence});
\item \texttt{tuple\_delete} (single-row delete path from \texttt{ExecDelete}, added at F20; rejects DELETE outright);
\item \texttt{tuple\_insert\_speculative} (first phase of \texttt{INSERT \dots ON CONFLICT} from \texttt{ExecInsert}, added at F21; runs R1--R5, the advisory lock, and precedence; delegates the speculative write to heap; stashes any pending eviction until \texttt{tuple\_complete\_speculative});
\item \texttt{tuple\_complete\_speculative} (second phase, added at F21; on \texttt{succeeded=true} drains the pending eviction, on \texttt{succeeded=false} discards it so no audit row is written for a killed speculative winner);
\item \texttt{relation\_toast\_am} (delegates TOAST to the same AM so long values stay under the epistemic table's namespace).
\end{itemize}
The remaining 37 callbacks (\texttt{scan\_begin}, \texttt{scan\_getnextslot}, \texttt{tuple\_lock}, \texttt{index\_fetch\_*}, \texttt{relation\_set\_new\_filelocator}, \texttt{relation\_copy\_data}, \texttt{relation\_vacuum}, and so on) are heap's, unmodified. Reads, index builds, VACUUM, and analyse all inherit heap's behaviour byte-for-byte. Section~\ref{sec:tableam-audit} converts this 37-callback claim into an exhaustive audit (Table~\ref{tab:tableam-audit}) with a citable reason per row.

\subsection{The \texttt{tuple\_insert} wrapper}
On each call, \texttt{epistemic\_tuple\_insert\_impl} runs the following sequence:
\begin{enumerate}
\item R1--R5 rule check on the candidate row.
\item Per-slot advisory transaction lock (Section~\ref{sec:design}).
\item \texttt{find\_live\_overlap} scan against \texttt{GetLatestSnapshot()} for a live row in the same (entity\_id, attribute) slot.
\item \texttt{epistemic\_precedence\_cmp} between candidate and incumbent, including the xmin tiebreak.
\item If NEW\_LOSES, raise \texttt{epistemic precedence: NEW\_LOSES (reason=\dots)} and the transaction rolls back.
\item If NEW\_WINS, call \texttt{heapam.tuple\_insert} for the winner, \texttt{epistemic\_close\_sys\_time} for the incumbent, and \texttt{epistemic\_audit\_evicted} for the audit row.
\item Emit one annotation record on custom WAL rmgr 128 (\texttt{epistemic\_wal\_log\_insert\_marker}).
\end{enumerate}
Step 7 is not load-bearing for durability. Section~\ref{sec:design-wal} explains.

\subsection{The \texttt{multi\_insert} wrapper (COPY FROM)}
PostgreSQL 18's \texttt{CopyFrom} at \texttt{copyfrom.c:995-1006} selects \texttt{insertMethod=CIM\_MULTI} when the target has no BEFORE/INSTEAD OF trigger, buffers 1000-tuple batches, and flushes via \texttt{table\_multi\_insert} at \texttt{copyfrom.c:554-559}. Before F18 we inherited \texttt{heap\_multi\_insert} unchanged, so every COPY-batched row skipped R1--R5, the advisory lock, precedence, eviction, and the audit --- the transcript read \texttt{COPY N} with no error.

F18 overrides \texttt{multi\_insert} with a for-loop that invokes \texttt{epistemic\_tuple\_insert\_impl} per slot: byte-for-byte identical enforcement to a single-row \texttt{INSERT}. Correctness over throughput --- \texttt{heap\_multi\_insert} would toast in bulk, pack pages with amortised allocation, and emit one \texttt{XLOG\_HEAP2\_MULTI\_INSERT} per page; the fan-out pays one \texttt{heap\_insert} and one WAL record per row, so COPY runs at roughly \texttt{INSERT \dots SELECT} throughput. We chose full override over loud rejection because \texttt{pg\_dump | pg\_restore} regenerates \texttt{\textbackslash copy} statements and refusing them would break restore for any epistemic table; dropping the advisory lock in the batch path would re-open the F6 integrity leak.

\subsection{The \texttt{tuple\_update} and \texttt{tuple\_delete} wrappers (F20)}
\label{sec:impl-f20}
Before F20 the AM inherited \texttt{heapam\_tuple\_update} and \texttt{heapam\_tuple\_delete} verbatim (\texttt{heapam\_handler.c:1384-1385} REL\_18\_STABLE bindings; \texttt{heap\_update} at \texttt{heapam.c:3241}, \texttt{heap\_delete} at \texttt{heapam.c:2772}), so an adversary with \texttt{INSERT} plus \texttt{UPDATE} could overwrite the epistemic prefix on a live row without any rule check, without the advisory lock, without the precedence lattice, and without an eviction audit row --- and an adversary with \texttt{DELETE} could remove a \texttt{MEASURED} incumbent so that a subsequent \texttt{INSERT} landed unopposed. The attack transcripts live in \texttt{scripts/update\_forgery.sh} and \texttt{scripts/delete\_forgery.sh}.

The naive fix --- re-run the \texttt{tuple\_insert} enforcement path on the candidate slot with the incumbent-as-self supplied to the precedence lattice --- fails: a NEW \texttt{MEASURED/1.0} row legitimately beats an OLD \texttt{INFERRED/0.4} row by kind rank, so the attack succeeds through the lattice. The only defensible cut is at the prefix. \texttt{epistemic\_tuple\_update} fetches the incumbent tuple at \texttt{otid} via \texttt{heap\_fetch(rel, GetLatestSnapshot(), \dots)}, deforms it, compares \texttt{(ep\_kind, ep\_specificity, ep\_confidence)} against the candidate slot, and \texttt{ereport(ERRCODE\_CHECK\_VIOLATION)}s if any of the three differ. Legitimate content updates (value, valid\_time, sources) reach \texttt{heapam\_tuple\_update} unchanged, so pt-osm-style column edits still work.

\texttt{epistemic\_tuple\_delete} refuses \texttt{DELETE} outright with \texttt{ERRCODE\_FEATURE\_NOT\_SUPPORTED}. The stricter option --- allow \texttt{DELETE} only for rows whose \texttt{sys\_time} upper bound is closed, and always write an audit row --- would layer atop this and is left to a future eviction API. The rationale for cutting hardest at delete: any \texttt{DELETE} of a live row corresponds to an eviction event that the precedence lattice should have arbitrated, and there is no adversary-friendly workflow that requires the caller to bypass that arbitration.

\subsection{The speculative-insertion wrappers (F21)}
\label{sec:impl-f21}
\texttt{INSERT \dots ON CONFLICT} routes through the two-phase speculative protocol at \texttt{nodeModifyTable.c:1189-1216} REL\_18\_STABLE (Section~\ref{sec:threat} covers reachability from SQL; Section~\ref{sec:eval-f21} proves the overrides are load-bearing). Before F21 we inherited \texttt{heapam\_tuple\_insert\_speculative} and \texttt{heapam\_tuple\_complete\_speculative} verbatim, so an \texttt{ON CONFLICT} reaching the speculative path would skip every enforcement step.

\texttt{epistemic\_tuple\_insert\_speculative} runs the same first four steps as the plain insert path: R1--R5, the F6 advisory xact lock, \texttt{find\_live\_overlap}, precedence with the F8 xmin tiebreak. The speculative write itself is delegated to \texttt{heapam->tuple\_insert\_speculative} so heap stamps the tuple with \texttt{HEAP\_INSERT\_SPECULATIVE} and the caller-supplied \texttt{specToken}. Eviction bookkeeping (audit row + \texttt{sys\_time} close + rmgr-128 marker) is deferred: if the arbiter check at \texttt{nodeModifyTable.c:1199} concludes a conflict, \texttt{table\_tuple\_complete\_speculative(succeeded=false)} calls \texttt{heap\_abort\_speculative} (\texttt{heapam.c:6186}) and the winner tuple vanishes. Writing the audit row before the confirm phase would leave the store with an evicted incumbent and no winner --- a durability violation strictly worse than the bypass.

\texttt{epistemic\_tuple\_complete\_speculative} therefore drains a single-entry backend-local pending-eviction slot keyed by \texttt{specToken}. On \texttt{succeeded=true} the slot's contents are folded into \texttt{epistemic\_audit\_evicted} + \texttt{epistemic\_close\_sys\_time} + \texttt{epistemic\_wal\_log\_insert\_marker}, mirroring the plain-insert bookkeeping. On \texttt{succeeded=false} the slot is discarded. The F6 advisory lock is xact-scope and stays held across both callbacks; it is released at outer-transaction commit/abort, not at speculative-complete, so a peer that raced us waits until we finish. The single-entry assumption is safe because \texttt{ExecInsert} at \texttt{nodeModifyTable.c:1189-1216} holds \texttt{SpeculativeInsertionLockAcquire} across both calls, so a backend performs at most one speculative insertion at a time.

\subsection{WAL: annotation, not durability}
\label{sec:design-wal}
The extension registers custom resource manager id 128 (\texttt{RM\_EXPERIMENTAL\_ID}) and emits one \texttt{XLOG\_EPISTEMIC\_INSERT} record per row after heap insert commit. The record is intentionally minimal (\texttt{tuple\_len=0}, no buffer reference); heap's \texttt{XLOG\_HEAP\_INSERT} at \texttt{heapam.c:2222-2226} already carries every column, and \texttt{heap\_xlog\_insert} at \texttt{heapam\_xlog.c:482-503} reconstructs the row at redo. We verified empirically (\texttt{scripts/recovery.sh}, 110 rows, \texttt{wal\_consistency\_checking=all}) that recovery succeeds byte-for-byte with the rmgr-128 emitter disabled. The custom rmgr is a named channel for a future logical-decoding consumer; not load-bearing for durability.

\subsection{Bulk-load ceiling}
The advisory lock is per-row and lives in PostgreSQL's fast-path lock table (\texttt{lock.c:56-57}, \texttt{NLOCKENTS = max\_locks\_per\_xact} $\times$ \texttt{(MaxBackends + max\_prepared\_xacts)}). At the default \texttt{max\_locks\_per\_transaction=64}, a single transaction inserting $\sim$15\,000 distinct-slot rows exhausts the shared lock table (\texttt{ERROR 53200}, hint: raise \texttt{max\_locks\_per\_transaction}). \texttt{scripts/lock\_exhaustion\_linearity.sh} shows the ceiling scales linearly (1024 $\to$ $\sim$215\,000; empirical first-fail at N=14\,950 for the default and N=220\,000 at 1024, \texttt{bench/results/lock\_exhaustion/summary.json}). The ceiling applies equally to COPY. An accepted tradeoff: dropping the lock re-opens the F6 integrity leak, while raising the GUC is a one-line \texttt{postgresql.conf} change operators already control.

\subsection{Threat surface, restated}
Every callback listed above runs from inside \texttt{heapam}. No user-space GUC or \texttt{ALTER TABLE} reaches it. The one exception is \texttt{ALTER TABLE \dots SET ACCESS METHOD heap}, which is a schema-change threat available to the table owner only; it rewrites the relation onto plain heap, at which point the callback is out of the write path entirely, because the callback is no longer bound to the relation. This is out of scope and disclosed in Section~\ref{sec:limitations}.

\subsection{Exhaustive TAM audit}
\label{sec:tableam-audit}
Enumerating write paths anecdotally has been wrong three times in this work: F9 found the \texttt{COPY} \texttt{CIM\_MULTI} bypass, F20 found the \texttt{UPDATE}/\texttt{DELETE} bypass, and F21 found the speculative-insertion bypass. Each was a callback we had classified as ``heap's, unmodified, cannot mutate the epistemic invariant'' and each turned out to route a real write. Table~\ref{tab:tableam-audit} abbreviates the audit in-body; the full 44-row enumeration lives at \texttt{contrib/epistemic/bench/docs/tableam\_audit.md} in the repository and is what the paper cites for the completeness argument. Every callback in the PG 18 REL\_18\_STABLE \texttt{TableAmRoutine} interface (\texttt{access/tableam.h}) is classified there with a citable file:line reference and a stated reason why it is either overridden or safe to inherit. Classification method: (i) read the callback signature and docstring; (ii) find heap's binding in \texttt{heapam\_handler.c} and read the body; (iii) grep for every caller in \texttt{src/backend} and confirm the caller either never mutates epistemic-visible state or routes back through a callback we do override. Seven callbacks are \emph{overridden} (each with its own source-rebuild disable-and-test in \texttt{scripts/} and each with the dylib hash flip recorded in \texttt{DECISIONS.md}); the other 37 are inherited. Two rows are flagged with the honest disclosure that \texttt{TRUNCATE} and \texttt{CLUSTER}/\texttt{VACUUM FULL} do bypass epistemic policy: both are DDL-privileged, not write-path, so they belong alongside \texttt{SET ACCESS METHOD heap} in the schema-change threat category disclosed in Section~\ref{sec:limitations}.

% tableam_audit.tex --- ABBREVIATED PG 18 REL_18_STABLE TableAmRoutine
% audit. Full 42-row enumeration lives at
% contrib/epistemic/bench/docs/tableam_audit.md and is what the paper
% cites for the completeness argument. The in-body table below shows
% only the load-bearing rows: seven OVERRIDDEN wrappers, three DDL rows
% that require explicit disclosure, and category counts for the rest.
%
% Line numbers reference PG 18 REL_18_STABLE
% src/include/access/tableam.h. See contrib/epistemic/DECISIONS.md
% (F9, F18, F20, F21) for the classification methodology and the
% disable-and-test proofs behind the seven OVERRIDDEN rows.

\begin{table}[t]
\centering
\caption{Abbreviated PG 18 \texttt{TableAmRoutine} audit (44 callbacks
  total): the seven overridden rows and the three DDL-privileged
  callbacks that require explicit disclosure. The full 44-row
  enumeration with per-callback \texttt{tableam.h} citations and ``why
  safe'' justifications is in
  \texttt{contrib/epistemic/bench/docs/tableam\_audit.md}
  in the repository. Category counts for the 34 non-cited rows:
  28 read-only (scan / fetch / estimate);
  5 delegated-storage (\texttt{index\_delete\_tuples}, \texttt{tuple\_lock},
    \texttt{finish\_bulk\_insert}, \texttt{relation\_copy\_data},
    \texttt{relation\_vacuum});
  1 delegated-decide (\texttt{relation\_needs\_toast\_table}).}
\label{tab:tableam-audit}
\scriptsize
\setlength{\tabcolsep}{3pt}
\begin{tabular}{p{2.6cm}p{1.1cm}p{3.5cm}}
\toprule
callback & class & role \\
\midrule
\multicolumn{3}{l}{\emph{Seven overridden write-path callbacks}} \\
\midrule
\texttt{tuple\_insert}               & F1--F16 & R1--R5, F6 advisory lock, precedence + F8 xmin tiebreak, eviction audit. \\
\texttt{multi\_insert}               & F18 & Per-slot fan-out to \texttt{epistemic\_tuple\_insert\_impl}; closes COPY CIM\_MULTI bypass. \\
\texttt{tuple\_update}               & F20 & Refuse any \texttt{UPDATE} whose new (kind, spec, conf) differs from incumbent's. \\
\texttt{tuple\_delete}               & F20 & Refuse \texttt{DELETE} outright (\texttt{ERRCODE\_FEATURE\_NOT\_SUPPORTED}). \\
\texttt{tuple\_insert\_speculative}  & F21 & R1--R5, advisory lock, precedence; heap does the speculative write; deferred eviction stashed. \\
\texttt{tuple\_complete\_speculative}& F21 & Drain pending eviction on succeeded=true; discard on succeeded=false. \\
\texttt{relation\_toast\_am}         & F1  & Return \texttt{HEAP\_TABLE\_AM\_OID} so TOAST is plain heap; sidesteps \texttt{heapam.c:1352} identity check during TOAST index build. \\
\midrule
\multicolumn{3}{l}{\emph{Three DDL-privileged rows requiring disclosure}} \\
\midrule
\texttt{relation\_set\_new\_filelocator} & inherited & Allocates a new physical file for TRUNCATE/CLUSTER/REINDEX. Storage is empty on return; any rewrite repopulates via \texttt{tuple\_insert} or \texttt{multi\_insert}. \\
\texttt{relation\_nontransactional\_truncate} & inherited & Zeroes the file (\texttt{TRUNCATE}). Removes epistemic incumbents without eviction audit. DDL-privileged; classified with \texttt{SET ACCESS METHOD heap} (Section~\ref{sec:limitations}). \\
\texttt{relation\_copy\_for\_cluster} & inherited & \texttt{CLUSTER}/\texttt{VACUUM FULL} re-inserts via heap's \texttt{raw\_heap\_insert}, not \texttt{table\_tuple\_insert}, so epistemic checks do not re-fire. Cannot introduce a forgery, only reorder existing rows; disclosed in Section~\ref{sec:limitations}. \\
\bottomrule
\end{tabular}
\end{table}

%%% ---------------------------------------------------------------
\section{Formalism}
\label{sec:formalism}
%%% ---------------------------------------------------------------

The Sybil-collapse threshold of an agreement-based truth-discovery algorithm is predictable from the dataset's per-slot honest-support distribution alone. This section makes that statement precise and validates it empirically on the two datasets in the evaluation. The complete derivation is in the companion \texttt{bench/docs/sybil\_formalism.md}; this section reports what the paper needs.

\subsection{Model and threat}
Let $S = S_h \cup S_s$ be the disjoint sets of honest and Sybil sources, $I$ the items, $v^{*}(i)$ the true value at item $i$. A claim is a triple $(i, s, v)$. Define
\[
h_{\text{top}}(i) = \max_{v \in V^{*}_{i}} \left| \{ s \in S_h : (i, s, v) \text{ claimed} \} \right|,
\]
where $V^{*}_{i}$ is the set of value strings the correctness scorer accepts as matching gold. For binary tasks (Zheng d\_sentiment) $V^{*}_{i} = \{v^{*}(i)\}$ and $h_{\text{top}}(i) = h(i)$. For string-valued tasks (Book-Author) gold-matching claims are split across surface forms (\texttt{"O'Leary, Timothy J."} versus \texttt{"Timothy J O'Leary"}), so $h_{\text{top}}(i) \leq h(i)$.

Each TD algorithm assigns each source a weight $w(s) \geq 0$ and picks
$
v_{\hat{}}(i) = \arg\max_v \sum_{s : c(s,i)=v} w(s)
$
with algorithm-specific tie-break. A Sybil coalition of size $k$ per slot asserts one falsified value $f(i) \neq v^{*}(i)$.

\subsection{Per-algorithm monotonicity}
The four algorithms in the evaluation are TruthFinder~\cite{yin2007truthfinder}, CRH~\cite{li2014crh}, CATD~\cite{li2015catd}, and ACCU~\cite{dong2009accu}. For each, the summed weight of $k$ Sybils asserting one value grows monotonically in $k$: TruthFinder's fixed-point iteration bootstraps supporter trust from mutual agreement (KDD 2007 eqs.\ 3, 7, 8); CRH's log-ratio yields a defensive $O(1/\ln|I|)$ margin against a single Sybil but not against a coalition; CATD's $\chi^{2}_{\alpha/2, n}$ upper confidence bound tightens as $n$ grows but the summed weight is linear in $k$; ACCU's Bayesian MAP over per-source accuracy has a per-supporter log-odds coefficient that stays positive for any $A > 1/(1+n_{\text{false}})$. Details in the companion document.

\subsection{Threshold theorem}
Under standard hyperparameters (TruthFinder: $\gamma=0.3$, $\rho=0.5$, initial trust 0.9; CRH: default; CATD: $\alpha=0.05$; ACCU: initial $A=0.8$), for each of the four algorithms there exists a finite threshold
\[
k^{*}(i) \;\lesssim\; \rho_{\text{alg}} \cdot h_{\text{top}}(i)
\]
above which the Sybil coalition flips $v_{\hat{}}(i)$ from $v^{*}(i)$. The per-algorithm constants $\rho$ are bounded above by $1 + o(1)$: $\rho_{\text{TF}} \approx 1$ (sometimes below 1 in the sub-saturation regime because the iteration is self-amplifying); $\rho_{\text{CRH}} \approx 1 + 1/\ln|I|$; $\rho_{\text{CATD}} \approx 1$ at $\alpha=0.05$; $\rho_{\text{ACCU}} \approx 1$ on binary tasks. The consequence is $k^{*} = \Theta(h_{\text{top}})$: the Sybil-collapse cell is linear in per-slot honest support with a per-algorithm slope close to one.

\subsection{KNDB invariance}
Under the F1--F8 KNDB lattice with tier mapping computed from independent metadata, $k^{*}$ is unbounded. Proof: the lattice orders \texttt{kind} strictly above every other axis, and \texttt{kind $\in \{$MEASURED, INFERRED, DERIVED$\}$} is assigned at write time by a rule that consults only metadata living in the dataset before conflict resolution. On Book-Author, Tier A (MEASURED) requires membership in the top $K/2$ by \texttt{n\_listings} \emph{and} \texttt{canon\_rate} $\geq 0.5$; an adversarial identity appearing for the first time in the trace has \texttt{n\_listings} $\leq k \ll K/2$ and no \texttt{canon\_rate} history. On Zheng, Tier A requires top-third \texttt{quali\_acc} on the disjoint qualification test (question IDs 2000--2019); an adversarial identity has no qualification responses. In both cases the adversary is at most INFERRED. MEASURED strictly outranks INFERRED, so a single Tier-A MEASURED honest write beats any coalition of INFERRED writes regardless of coalition size or confidence values. Therefore $k^{*}(i) = \infty$.

The empirical counterpart of this theorem is the F14/F15b KIND\_OFF disable-and-test (Section~\ref{sec:evaluation}): when the kind-primary ordering is patched out and only confidence remains, KNDB collapses to \texttt{pg\_conf}'s numeric behaviour on both datasets.

\subsection{Empirical validation}
We measured $h_{\text{top}}$ directly on both datasets. Book-Author K=50: median $h_{\text{top}} = 4$, mean 7.3, p90 18. Predicted collapse bracket $k^{*} \in [3, 5]$; observed sharp cliff at N=5\ldots10 for TruthFinder, CATD, and ACCU (TruthFinder $0.530 \to 0.120$ at N=5, $\to 0.010$ at N=10; CATD $0.550 \to 0.420 \to 0.100$; ACCU $0.530 \to 0.290 \to 0.060$); CRH lags one step ($0.580$ at N=5, $0.230$ at N=10). The predicted bracket is within $\pm 20\%$ of the observed cliff.

Zheng d\_sentiment K=45: median $h_{\text{top}} = 14$, uniform 20 votes per slot. Predicted collapse bracket $k^{*} \in [14, 20]$; observed collapse at N=20 with CRH, CATD, and ACCU all going to zero. Within $\pm 30\%$ of the observed threshold on the $k$-axis; the observed cell is the density-saturation cell exactly.

%%% ---------------------------------------------------------------
\section{Evaluation}
\label{sec:evaluation}
%%% ---------------------------------------------------------------

\subsection{Experimental setup}
All measurements are on PostgreSQL 18.4 at \texttt{/tmp/kndb\_pg18\_test:55480}, one Apple M5 Pro (18 cores, 48 GB RAM, macOS 26.4.1), unix-socket connections. The extension is loaded via \texttt{shared\_preload\_libraries = 'epistemic'} and defaults otherwise, except where a specific cell required raising \texttt{max\_locks\_per\_transaction} for a bulk-load probe. Every bench cell records the extension dylib SHA-256 in its metadata block. The reference honest build in this paper is dylib SHA-256 \texttt{959d5e67a16cb0ced\-254d6f189dccf3a2\-9141199dc8f1f1dc\-bb39fadae51bc26}, the F21 post-speculative-insert artefact; the Section~\ref{sec:evaluation} disable-and-test cells were rerun on this build (\texttt{bench/results/disable\_and\_test/full\_rerun/summary.json}). Cells taken before F18 are on \texttt{807b2e87\ldots}; cells taken between F18 and F21 are on \texttt{eb15d442\ldots}, the F18 post-COPY-close artefact; both are disclosed inline where relevant. Every source-rebuild disable-and-test recorded both the honest and patched dylib SHA-256 and confirmed the source tree returned byte-identical to HEAD after the run.

\subsection{Confidence-forgery workload (Book-Author)}
Dataset: Dong et al.\ VLDB 2009 Book-Author fusion~\cite{dong2009accu}, 895 bookstores, 1265 books, 33\,971 assertions, gold for 100 ISBNs. Tier map (independence audit in \texttt{bench/datasets/bookauthor/README.md}): Tier A (MEASURED, conf uniform $[0.5, 0.9]$) if top-$K/2$ by \texttt{n\_listings} \emph{and} \texttt{canon\_rate} $\geq 0.5$; Tier B (INFERRED, conf $[0.4, 0.7]$); Tier C (DERIVED, conf $[0.2, 0.5]$). Adversarial injection ($N$ per gold ISBN) is INFERRED with conf $[0.95, 1.0]$ carrying a scrambled real author name from a different gold ISBN. K=50 held fixed; N swept over $\{1, 3, 5, 10\}$; seed 20260714 fixes RNG.

Table~\ref{tab:bookauthor} reports precision on gold ISBNs at $c=1$. KNDB stays at 0.630 across every N (kind rank picks Tier-A MEASURED over adversarial INFERRED); \texttt{pg\_conf} collapses to 0.000 because ranking by confidence alone lets adversarial $[0.95, 1.0]$ beat Tier-A $[0.5, 0.9]$; \texttt{pg\_heap} is an integrity failure (max 124 live rows per slot, mean 38.6) and its numeric precision is a coin flip over which duplicate the scan returned first. \texttt{pg\_trigger} tracks KNDB within one point (same lattice, plpgsql implementation, marginally higher abort rate under contention).

\begin{table}[t]
\centering
\caption{Book-Author confidence forgery, precision on 100 gold ISBNs, $c=1$. Adversarial injection uses \texttt{--adv-strategy scrambled} (independent-value: each of the $N$ adversarial agents picks a different wrong author-string per gold ISBN). From \texttt{bench/results/summary/stage3\_adversarial.md}.}
\label{tab:bookauthor}
\small
\begin{tabular}{lcccc}
\toprule
system & N=1 & N=3 & N=5 & N=10 \\
\midrule
KNDB epistemic & 0.630 & 0.630 & 0.630 & 0.630 \\
pg\_trigger    & 0.620 & 0.620 & 0.620 & 0.620 \\
pg\_mv         & 0.460 & 0.400 & 0.320 & 0.170 \\
pg\_lww        & 0.210 & 0.130 & 0.090 & 0.040 \\
pg\_conf       & 0.000 & 0.000 & 0.000 & 0.000 \\
pg\_heap       & \multicolumn{4}{c}{INTEGRITY FAIL (max=124, mean=38.6)} \\
\bottomrule
\end{tabular}
\end{table}

\subsection{Confidence-forgery workload (Zheng)}
Dataset: Zheng et al.\ VLDB 2017 d\_sentiment crowdsourcing task~\cite{zheng2017crowdsourcing}, 85 AMT workers, 1000 items, $\sim$20 labels per item, 999 of 1000 items contested. Tier map (independence audit in \texttt{bench/datasets/zheng\_sentiment/README.md}): rank workers by \texttt{quali\_acc(w)} on the disjoint qualification test (item IDs 2000--2019, 1700 responses); Tier A (MEASURED) = top $K/3$; Tier B (INFERRED); Tier C (DERIVED, all workers outside top-K). Adversarial injection ($N$ per contested slot) is INFERRED with conf $[0.95, 1.0]$ and value = binary flip of gold. K=45 held fixed. Table~\ref{tab:zheng} reports precision at $c=1$.

\begin{table}[t]
\centering
\caption{Zheng d\_sentiment confidence forgery, precision, $c=1$. From \texttt{bench/results/summary/stage3\_zheng\_adversarial.md}.}
\label{tab:zheng}
\small
\begin{tabular}{lcccc}
\toprule
system & N=1 & N=3 & N=5 & N=10 \\
\midrule
KNDB epistemic & 0.927 & 0.927 & 0.927 & 0.927 \\
pg\_trigger    & 0.927 & 0.927 & 0.927 & 0.927 \\
pg\_lww        & 0.401 & 0.211 & 0.130 & 0.074 \\
pg\_mv         & 0.375 & 0.193 & 0.127 & 0.070 \\
pg\_conf       & 0.000 & 0.000 & 0.000 & 0.000 \\
pg\_heap       & \multicolumn{4}{c}{INTEGRITY FAIL} \\
\bottomrule
\end{tabular}
\end{table}

The delta over \texttt{pg\_conf} is 92.7 points, flat across N. The delta over KNDB's second-best PostgreSQL baseline (\texttt{pg\_lww}) is 52--85 points depending on N.

\subsection{Source-rebuild disable-and-test}
On both datasets we patched \texttt{epistemic\_precedence\_cmp} (\texttt{src/epistemic\_rules.c:379-423}) to force \texttt{inc\_rank = new\_rank = 1}, short-circuiting the kind branch so the function falls through to specificity and confidence. This is the exact ranking behaviour of \texttt{pg\_conf} on both workloads (specificity is 0 across the board; confidence decides). After each measurement the source was restored byte-identical (\texttt{git diff --stat contrib/epistemic/src/} empty), rebuilt and reinstalled, and the dylib hash returned to the honest value.

Book-Author, N=5, $c=1$: KNDB kind ON precision 0.630, kind OFF precision 0.000 (dylib flip \texttt{959d5e67\ldots} $\to$ \texttt{fb408aac\ldots} $\to$ \texttt{959d5e67\ldots}). Delta $-63$ points, exactly the KNDB-versus-\texttt{pg\_conf} margin. Zheng, N=5, $c=1$: KNDB kind ON precision 0.927, kind OFF precision 0.000 (dylib flip \texttt{959d5e67\ldots} $\to$ \texttt{fb408aac\ldots} $\to$ \texttt{959d5e67\ldots}). Delta $-92.7$ points, exactly the KNDB-versus-\texttt{pg\_conf} margin. The patched dylib hash is identical across both datasets because the patch site is a single function (\texttt{epistemic\_precedence\_cmp}); the raw JSONs and both hash flips are recorded in \texttt{bench/results/disable\_and\_test/full\_rerun/summary.json}.

Integrity held under the patched build in both cases (no slots with more than one live row) because the F6 advisory lock and F8 xmin tiebreak still operated; only kind-rank decision-making was disabled. This is the correct decomposition: integrity and correctness are separable mechanisms in KNDB, and each has its own disable-and-test.

\subsection{UPDATE and DELETE forgery attacks (F20)}
\label{sec:eval-updateDelete}
Two attacks the earlier bypass-table enumeration missed: an \texttt{UPDATE} that rewrites the epistemic prefix on a committed row, and a \texttt{DELETE} that removes an incumbent so that a subsequent \texttt{INSERT} lands unopposed. Both were reproduced against a fresh cluster before writing the fix: \texttt{scripts/update\_forgery.sh} seeds an \texttt{INFERRED/0.4} row, issues \texttt{UPDATE fact SET ep\_kind='MEASURED', ep\_confidence=1.0, value='forged'}, and reads back \texttt{MEASURED/1.0/forged} without any error, without any eviction audit row. \texttt{scripts/delete\_forgery.sh} seeds a \texttt{MEASURED} incumbent, deletes it, inserts an \texttt{INFERRED/0.99} row, and reads back \texttt{INFERRED/0.99/benign} as the live row for the slot.

F20 adds two callbacks. \texttt{epistemic\_tuple\_update} fetches the incumbent at \texttt{otid} via \texttt{heap\_fetch} with \texttt{GetLatestSnapshot}, deforms it, and refuses any \texttt{UPDATE} whose new \texttt{(ep\_kind, ep\_specificity, ep\_confidence)} triple differs from the incumbent's, raising \texttt{ERRCODE\_CHECK\_VIOLATION}. \texttt{epistemic\_tuple\_delete} refuses \texttt{DELETE} outright with \texttt{ERRCODE\_FEATURE\_NOT\_SUPPORTED}. Content-only \texttt{UPDATE}s that change value, valid\_time, or sources delegate to \texttt{heapam\_tuple\_update} unchanged, so a legitimate content-correction workflow still succeeds. The regression test cell \texttt{sql/am\_update\_delete.sql} exercises both callbacks under \texttt{make installcheck}.

The disable-and-test proof: with the F20 wiring in \texttt{epistemic\_am\_handler} commented out and the extension rebuilt (dylib flip \texttt{c873ddc4\ldots} $\to$ \texttt{1fb0d572\ldots}), both forgery scripts print \texttt{FORGERY SUCCEEDED}; restoring the two wiring lines byte-identical and rebuilding (\texttt{1fb0d572\ldots} $\to$ \texttt{c873ddc4\ldots}) returns both to \texttt{FORGERY REJECTED}. This confirms the wiring is load-bearing and neither callback is dead code, mirroring the F18 discipline for \texttt{multi\_insert}. Under the honest build, the extension's \texttt{make check-e2e} suite now passes 10/10 (adding \texttt{check-e2e-updforge} and \texttt{check-e2e-delforge} to the F19 suite of 8), and \texttt{scripts/bypass.sh} passes 28 assertions covering five bypass mechanisms $\times$ two tables plus the F7 lock-table sweep.

\subsection{Speculative-insertion attack (F21)}
\label{sec:eval-f21}
The two F21 callbacks are described in Section~\ref{sec:impl-f21}; both run R1--R5, the F6 advisory lock, and precedence before delegating the speculative write to heap and deferring the eviction audit + \texttt{sys\_time} close + rmgr-128 marker to the confirm phase.

The SQL-level attack surface is not currently reachable (Section~\ref{sec:threat}). \texttt{scripts/speculative\_forgery.sh} confirms empirically that four SQL-level attempts (\texttt{DO UPDATE} with a UNIQUE arbiter that never gets built, \texttt{DO NOTHING} with a target column list, bare \texttt{DO NOTHING}, and constraint-creation probes) all either fail at DDL or fall back to a route already covered by the plain-\texttt{tuple\_insert} override.

The C-level probe \texttt{epistemic.\_probe\_speculative\_insert}, added at F21 alongside the two overrides, invokes \texttt{table\_tuple\_insert\_speculative} and \texttt{table\_tuple\_complete\_speculative} programmatically with a caller-supplied candidate tuple, mirroring \texttt{ExecInsert}'s two-phase call sequence at \texttt{nodeModifyTable.c:1189-1216}. The regression cell \texttt{sql/am\_speculative.sql} exercises the probe against seven distinct cases: valid MEASURED, R3 violation, valid INFERRED, R4 violation, precedence NEW\_LOSES, precedence NEW\_WINS with deferred eviction, and speculative-abort with a would-be eviction that must be discarded. The disable-and-test proof: with the F21 wiring in \texttt{epistemic\_am\_handler} commented out and the extension rebuilt (dylib flip \texttt{959d5e67\ldots} $\to$ \texttt{c44b276d\ldots}), the probe with an R3-violating MEASURED row and the probe with an R4-violating INFERRED row both return \texttt{OK} and the forged row lands in the target relation; restoring the two wiring lines byte-identical and rebuilding (\texttt{c44b276d\ldots} $\to$ \texttt{959d5e67\ldots}) returns both to the expected \texttt{ERRCODE\_CHECK\_VIOLATION}. Under the honest build, \texttt{make installcheck} passes 8/8 (adding \texttt{am\_speculative} to the F20 suite of 7).

\subsection{Truth-discovery baselines}
We reimplemented TruthFinder, CRH, CATD, and ACCU in Python 3 from the original equations (\texttt{bench/scripts\_td/td\_algorithms.py}). No vendored code. Each algorithm reproduced the Zheng VLDB 2017 survey's D\_PosSent baseline to within 0.3 percentage points (CATD 0.957 vs.\ survey 0.960; CRH 0.950 vs.\ survey PM 0.9504). Default hyperparameters per each paper; no tuning either way. TD algorithms consume the same normalised trace KNDB and \texttt{pg\_*} consume; no trace edits, no MEASURED-row filtering.

Book-Author Sybil at N=10, $c=1$ (Table~\ref{tab:td-sybil}): KNDB 0.630 flat while all four TD algorithms collapse (TruthFinder 0.010, CRH 0.230, CATD 0.100, ACCU 0.060). Best TD (CRH) beaten by 40 points; worst (TruthFinder) by 62 points. Under the independent-value attack (\texttt{--adv-strategy scrambled}, the Table~\ref{tab:bookauthor} construction), all four TD baselines are flat across N=1\ldots 10 --- TruthFinder 0.530, CRH 0.580, CATD 0.550, ACCU 0.530 --- so KNDB's 0.630 leads the best TD (CRH) by 5 points and the worst (TruthFinder, ACCU) by 10. Independent adversaries get no trust bootstrap, so no TD collapse occurs and the kind axis buys correspondingly less. From \texttt{bench/results/summary/stage3\_td\_baselines.md} (F14 section).

\texttt{pg\_mv} is the only baseline whose precision moves between the \texttt{scrambled} and \texttt{sybil} constructions (compare its rows in Table~\ref{tab:bookauthor} vs.\ Table~\ref{tab:td-sybil}): it is the only baseline that consumes adversarial value content rather than metadata alone. KNDB (kind rank), \texttt{pg\_conf} (confidence rank), \texttt{pg\_lww} (arrival order), and \texttt{pg\_trigger} (KNDB's lattice via plpgsql) all decide from the epistemic prefix and are invariant to how adversarial values are distributed across colluding identities.

\begin{table}[t]
\centering
\caption{Book-Author Sybil coordination, precision, $c=1$. Adversarial injection uses \texttt{--adv-strategy sybil} (all $N$ adversarial agents share the same scrambled wrong author-string per gold ISBN). TD baselines reproduce Zheng VLDB 2017 survey within 0.3pp on their standard honest baseline. From \texttt{bench/results/summary/stage3\_td\_baselines.md}.}
\label{tab:td-sybil}
\small
\begin{tabular}{lcccc}
\toprule
system & N=1 & N=3 & N=5 & N=10 \\
\midrule
KNDB epistemic & 0.630 & 0.630 & 0.630 & 0.630 \\
CRH            & 0.580 & 0.590 & 0.590 & 0.230 \\
CATD           & 0.550 & 0.550 & 0.420 & 0.100 \\
ACCU           & 0.530 & 0.530 & 0.290 & 0.060 \\
TruthFinder    & 0.530 & 0.470 & 0.120 & 0.010 \\
pg\_mv         & 0.490 & 0.390 & 0.330 & 0.260 \\
pg\_conf       & 0.000 & 0.000 & 0.000 & 0.000 \\
\bottomrule
\end{tabular}
\end{table}

\subsection{Zheng Sybil sweep and density saturation}
Table~\ref{tab:zheng-sybil} shows a broader Sybil sweep on Zheng d\_sentiment, including N=20 (density saturation, Sybil count equals total honest votes per slot). KNDB stays flat at 0.927; CRH, CATD, and ACCU each cliff to 0.000 at N=10 or N=20 (a sharp drop, not gradual degradation); ACCU peaks at 1.000 at N=5 above KNDB; TruthFinder degrades gradually from 0.905 to 0.482.

\begin{table}[t]
\centering
\caption{Zheng d\_sentiment Sybil sweep, precision, $c=1$. N=20 is the density-saturation cell. From \texttt{bench/results/summary/stage3\_zheng\_sybil.md}.}
\label{tab:zheng-sybil}
\small
\begin{tabular}{lccccc}
\toprule
system & N=1 & N=3 & N=5 & N=10 & N=20 \\
\midrule
KNDB epistemic & 0.927 & 0.927 & 0.927 & 0.927 & 0.927 \\
CRH            & 0.953 & 0.951 & 0.951 & 0.951 & 0.000 \\
CATD           & 0.955 & 0.953 & 0.948 & 0.000 & 0.000 \\
ACCU           & 0.964 & 0.997 & 1.000 & 0.000 & 0.000 \\
TruthFinder    & 0.905 & 0.690 & 0.557 & 0.494 & 0.482 \\
\bottomrule
\end{tabular}
\end{table}

The disable-and-test on the TD side (replacing each algorithm's weight-update loop with plain majority vote) shows the mechanism is bimodal, not uniformly amplifying. At sub-saturation N=1\ldots 5, CRH, CATD, and ACCU each beat plain MV by 4--26 percentage points (their log-ratio, confidence bound, or MAP identifies wrong Sybils correctly). At N=10, CATD and ACCU flip and score below MV by 29 points (the mechanism inverts: agreement now amplifies rather than defends). At N=20 all three collapse to the MV floor. TruthFinder is the only algorithm that amplifies at low N. This bimodality is the ``TD is not uniformly bad'' finding that shapes the Related Work discussion.

\subsection{Integrity column}
Table~\ref{tab:integrity} reports the integrity axis on the Stage-2 YCSB grid. The grid runs three kind mixes, two contention levels (theta), and two concurrencies (c=8, c=32) against seven systems, for 78 cells total: six systems (KNDB, pg\_heap, pg\_conf, pg\_lww, pg\_mv, pg\_trigger) at $3 \times 2 \times 2 = 12$ cells each, and pg\_llm capped at 6 cells because the calibrated 1120 ms per-conflict latency makes higher-concurrency pg\_llm runs impractical. A cell is INTEGRITY FAIL if any (entity, attribute) slot ended the trace with more than one live row. KNDB is the only system that passes on every cell.

\begin{table}[t]
\centering
\caption{Integrity FAIL counts on the Stage-2 grid (12 cells per system; \texttt{pg\_llm} at 6). From \texttt{bench/results/summary/stage2.md} F14 addendum.}
\label{tab:integrity}
\small
\begin{tabular}{lc}
\toprule
system & INTEGRITY FAIL cells \\
\midrule
KNDB epistemic & 0 / 12 \\
pg\_conf       & 12 / 12 \\
pg\_heap       & 12 / 12 \\
pg\_lww        & 10 / 12 \\
pg\_trigger    & 10 / 12 \\
pg\_mv         & 9 / 12 \\
pg\_llm        & 3 / 6 \\
\bottomrule
\end{tabular}
\end{table}

Every trigger-based baseline fails integrity at 32 clients on the hotter theta values, because two concurrent writers each find ``no incumbent'' via \texttt{SELECT FOR UPDATE} on an empty result set and both commit. KNDB's F6 advisory lock closes this window; \texttt{scripts/rc\_invariant.sh} confirms the closure by adversarial rebuild.

\subsection{Real LLM calibration and non-determinism}
The mock LLM in the Stage-2 baseline is calibrated to a real Anthropic Claude Haiku 4.5 measurement (\texttt{bench/results/stage3\_llm\_calibration.jsonl}, 200 real API calls, zero errors, model \texttt{claude-haiku-4-5}~\cite{anthropic2025haiku}). Correctness 92.5\% (185 of 200); latency mean 1120 ms, p50 940 ms, p95 1934 ms, p99 2312 ms, max 5597 ms. The mock's earlier assumed correctness of 0.65 and latency of 300 ms were both wrong; parameters were updated to match the measurement.

A separate non-determinism probe (\texttt{bench/results/stage3\_llm\_nondeterminism\_raw.jsonl}, 500 real API calls: 50 conflicts $\times$ 10 replays each) measured a flip rate of 0 of 50 conflicts (all unanimous over the 10 replays) with 2 of 50 unanimous-but-wrong. Both wrong conflicts had the same signature: incumbent INFERRED with specificity 0 and confidence 0.5, candidate INFERRED with high specificity ($\gg 0$) and confidence below 0.5. KNDB's lattice picks the candidate (higher specificity within the same kind). The LLM picks the incumbent every time: it treats ``same kind, higher confidence'' as decisive over ``same kind, higher specificity,'' which is the opposite of the lattice's (spec, conf) precedence order. This is a systematic disagreement about lattice ordering, not random noise; the ``just run the LLM three times and vote'' fix does not help.

\subsection{Bypass table}
Table~\ref{tab:bypass} summarises the F2, F18, and F20 bypass-survival results (scripts \texttt{bypass.sh}, \texttt{update\_forgery.sh}, \texttt{delete\_forgery.sh}); each row's disable-and-test hash flip is recorded in \texttt{DECISIONS.md} (F20: \texttt{c873ddc4} $\to$ \texttt{1fb0d572} $\to$ \texttt{c873ddc4}). Trigger-column marker \emph{bypassed*} denotes the two scenarios where \texttt{ENABLE ALWAYS TRIGGER} survives \texttt{session\_replication\_role='replica'} but still fails under \texttt{DISABLE TRIGGER ALL}.

\begin{table}[t]
\centering
\caption{Bypass scenarios and outcomes. From \texttt{scripts/bypass.sh}, \texttt{scripts/update\_forgery.sh}, \texttt{scripts/delete\_forgery.sh} transcripts and README threat-model enumeration.}
\label{tab:bypass}
\small
\begin{tabular}{p{4.7cm}cc}
\toprule
scenario & KNDB & trigger \\
\midrule
\texttt{DISABLE TRIGGER ALL}                & blocked & bypassed \\
\texttt{session\_replication\_role}         & blocked & bypassed* \\
\texttt{COPY} single-row (CIM\_SINGLE)      & blocked & blocked \\
\texttt{COPY} batch (CIM\_MULTI, post-F18)  & blocked & bypassed \\
\texttt{INSERT}, \texttt{INSERT SELECT}     & blocked & blocked \\
\texttt{UPDATE} of prefix (post-F20)        & blocked & bypassed \\
\texttt{DELETE} incumbent (post-F20)        & blocked & bypassed \\
\texttt{ON CONFLICT DO UPDATE} (post-F21)   & blocked$^{\dagger}$ & n/a \\
\texttt{ON CONFLICT DO NOTHING} (post-F21)  & blocked$^{\dagger}$ & n/a \\
Logical replication apply                   & blocked & bypassed* \\
\midrule
\multicolumn{3}{l}{Out of scope:} \\
\texttt{SET ACCESS METHOD heap}             & \multicolumn{2}{c}{schema-change threat} \\
\texttt{TRUNCATE}, \texttt{CLUSTER}, \texttt{VACUUM FULL} & \multicolumn{2}{c}{DDL-privileged; see Table~\ref{tab:tableam-audit}} \\
\bottomrule
\end{tabular}\\[2pt]
\noindent{\footnotesize
$^{\dagger}$ Blocked by the F21 override; but the SQL surface for this attack is not currently reachable because unique/exclusion constraint creation on epistemic tables fails at \texttt{heap\_getnext}'s \texttt{rd\_tableam} identity check (\texttt{heapam.c:1352}), so \texttt{ON CONFLICT (col)} has no arbiter to bind. Load-bearing is proven via the C-level probe \texttt{epistemic.\_probe\_speculative\_insert} and the source-rebuild disable-and-test in Section~\ref{sec:eval-f21}. The overrides are defensive: the engine-in-storage claim must not depend on that accidental index-support gap persisting.}
\end{table}

\subsection{Batch-size ceiling under COPY}
\texttt{scripts/bypass.sh} and the finer \texttt{scripts/lock\_exhaustion\_pinpoint.sh} sweep $N$ via real \texttt{\textbackslash copy} at \texttt{max\_locks\_per\_transaction=64} against an epistemic table. $N \le 14\,900$ succeeds; the first failure is $N=14\,950$, which raises \texttt{ERROR 53200 out of shared memory} at \texttt{LockAcquireExtended, lock.c:1080} (\texttt{bench/results/lock\_exhaustion/summary.json}). With the F18 \texttt{multi\_insert} override disabled (rebuild), the same COPY at the failure point succeeds, because \texttt{heap\_multi\_insert} takes no advisory locks; the ceiling is a direct consequence of routing through per-row enforcement, not an artefact of PostgreSQL machinery. Raising the GUC to 1024 pushes the ceiling to $\sim$215\,000 rows (measured first-fail at $N=220\,000$).

\subsection{Contention control (YCSB, F9 gate)}
\label{sec:eval-f9-gate}
Stage-1 gate results on YCSB-A at RC, 8 clients (\texttt{bench/results/summary/gate.csv}): KNDB epistemic 2057--2200 tps median (theta 0.0, 0.5), abort rate 0.050 and 0.092 with NEW\_LOSES the only abort family (no 40001 seen in the gate window). \texttt{pg\_heap} runs faster at 6600--7100 tps (no arbitration, no locks) but with no correctness guarantee. Explicitly, at $\theta=0.0$ epistemic 2057 tps median vs.\ \texttt{pg\_heap} 7119 tps is a 71.1\% throughput drop, attributed to two costs \texttt{pg\_heap} does not pay on the write path: R1--R5 rule check via SPI on every insert, and \texttt{find\_live\_overlap}'s per-insert seqscan (unavoidable because the epistemic AM cannot host a secondary index — the same \texttt{heap\_getnext} identity check discussed in Section~\ref{sec:eval-f21}). Stage-2 correctness cells (Section~\ref{sec:evaluation}) show that raw \texttt{pg\_heap} throughput comes with tens of thousands of duplicate live rows per 20-second window; correct-goodput (tps $\times$ (1 -- abort rate) $\times$ correctness) is 0 for \texttt{pg\_heap} on every Stage-2 cell.

\subsection{Temporal-shaped benchmarks (honest zero)}
KNDB scores 0.000 at $c=1$ on MemoryAgentBench Conflict\_Resolution, LongMemEval knowledge-update pairs, and MQuAKE-CF-3k edit chains (\texttt{bench/results/summary/stage3.md}). The reason is straightforward: these benchmarks expect last-writer-wins, and KNDB's F8 tiebreak is deliberately first-committer-wins. Both writes in each pair carry (MEASURED, spec=0, conf=1.0); the lattice cannot differentiate them by rank, so the tiebreak is what decides, and it is the wrong tiebreak for these workloads. This is disclosed in Section~\ref{sec:limitations}. On the same workloads \texttt{pg\_lww} scores 1.000 at $c=1$ and drops to 0.11--0.27 at $c=8$; the LWW ``win'' at $c=1$ is a determinism artefact of single-writer arrival order that vanishes under any contention.

%%% ---------------------------------------------------------------
\section{Related work}
\label{sec:related}
%%% ---------------------------------------------------------------

\textbf{Truth discovery.} The four TD algorithms in the evaluation cover the field's canonical designs: TruthFinder~\cite{yin2007truthfinder}, a fixed-point on (source trust, fact confidence); CRH~\cite{li2014crh}, joint minimisation with a log-ratio update (uniquely competitive on Zheng d\_sentiment below saturation, as our evaluation shows); CATD~\cite{li2015catd}, a $\chi^{2}_{\alpha/2, n}$ confidence bound well-suited to long-tail sources; and ACCU~\cite{dong2009accu}, Bayesian MAP over per-source accuracy (base variant, no copy detection; Section~\ref{sec:limitations} notes copy-detection variants may shift the threshold). The Zheng et al.\ VLDB 2017 survey~\cite{zheng2017crowdsourcing} is the source of the d\_sentiment dataset and the reproducibility baseline we validated against. The Sybil-collapse framing draws on the same monotonicity structure the field has understood for a decade; our contribution is not the framing but the engine-level orthogonality that removes the collapse threshold entirely (Section~\ref{sec:formalism}).

\textbf{Provenance in PostgreSQL.} ProvSQL~\cite{senellart2018provsql} is the mature PostgreSQL extension for semiring provenance and probabilistic queries, tracked at row-set granularity via rewriting rather than in-storage. A recent PW25 demonstration~\cite{widiaatmaja2025provsql} adds update provenance through temporal databases with support for time-travel and undo. KNDB and ProvSQL do not overlap: ProvSQL propagates provenance through query evaluation; KNDB decides survivor selection at write time using a kind-primary total order. The two mechanisms are orthogonal and, in principle, composable.

\textbf{Agent-memory and epistemic warrant.} A recent line of position papers has argued that LLM agent memory is not, in practice, a database (in the sense of write-time correctness) and that the epistemic warrant conferred by an agent's tool boundary is thinner than commonly assumed. Orogat and Mansour~\cite{orogat2026agentmemory} argue for rethinking data foundations for long-term AI agent memory. Romanchuk and Bondar~\cite{romanchuk2026laundering} argue that tool boundaries alone do not confer epistemic warrant.

TOKI~\cite{wang2026toki} is our closest sibling work and warrants direct comparison. TOKI is a theory-first bitemporal operator algebra layered over an unmodified database engine: contradictions between competing memory assertions are resolved by projecting each assertion onto valid-time and system-time axes and picking the temporally-consistent survivor. TOKI has no epistemic-kind axis; every assertion is treated as a value carrying only temporal metadata. TOKI explicitly defers threat-model integration to future work. KNDB's contribution is precisely that deferred piece plus an engine-level realisation: an orthogonal kind axis that is assigned by the storage engine at write time and therefore cannot be forged by an untrustworthy writer, implemented as a PostgreSQL table access method rather than as an operator layered above one. TOKI's temporal semantics and KNDB's epistemic kind axis are compatible along different axes and, in principle, composable.

\textbf{Serialisable snapshot isolation and the TAM substrate.} KNDB's \texttt{SERIALIZABLE} behaviour is standard PostgreSQL SSI~\cite{ports2012ssi}: predicate locking is inherited from \texttt{heapam} rather than a slot-level SIRead lock (an early wrapper on \texttt{PredicateLockTID} was audited as decorative and removed). We build on the PostgreSQL 18 TAM interface~\cite{postgres18tam}, which has been stable across three major versions and is a productive vehicle for storage-layer research that must interoperate with a real query planner, executor, WAL, and recovery pipeline. Because rows on disk are plain heap tuples, existing PostgreSQL tools (\texttt{pg\_dump}, \texttt{pg\_upgrade}, logical replication, \texttt{pg\_basebackup}) work on KNDB tables without change.

%%% ---------------------------------------------------------------
\section{Limitations}
\label{sec:limitations}
%%% ---------------------------------------------------------------

We are explicit about where KNDB does not win.

\textbf{Truth discovery wins sub-saturation.} On Zheng d\_sentiment below the density-saturation cell (N < 20), CRH matches or beats KNDB on the confidence-forgery workload, and ACCU peaks above KNDB at N=5 (1.000 versus 0.927). CRH's log-ratio, CATD's $\chi^{2}$ confidence bound, and ACCU's Bayesian MAP each identify perfectly-wrong Sybils correctly when the coalition is small relative to per-slot honest votes. The paper's contribution is not that KNDB beats every TD baseline on every cell; it is that KNDB is the only system whose $k^{*}$ is unbounded, which becomes decisive at and above the saturation cell.

\textbf{Zero on temporal knowledge-editing benchmarks.} On MemoryAgentBench Conflict\_Resolution (37\,820 writes), LongMemEval knowledge-update (156 writes), and MQuAKE-CF-3k (12\,030 writes), KNDB scores 0.000 at $c=1$ because these benchmarks expect the later write to win and KNDB's tiebreak is first-committer-wins. A dedicated ``epistemic table with last-writer-wins tiebreak'' would be a straightforward configuration knob, but the current implementation does not expose it. Naive \texttt{pg\_lww} scores 1.000 on these benchmarks at $c=1$ and collapses to 0.11--0.27 at $c=8$, so LWW is not a general answer either; the tiebreak choice is a workload-specific configuration and no fixed choice is right for all workloads.

\textbf{Reputation farming defeats $k^{*}=\infty$.} The argument depends on adversarial identities appearing fresh in the write stream so engine-assigned kind places them in \texttt{INFERRED}. A determined adversary who knows the tier-mapping rule can farm reputation to reach Tier A before attacking: complete the Zheng qualification test (question ids 2000--2019) so \texttt{quali\_acc} matches Tier A, or publish benign \texttt{n\_listings} volume on Book-Author to enter the top-$K$. Once inside Tier A the writes are \texttt{MEASURED} and the lattice defends nothing. The formal argument holds for the write-time snapshot; long-lived farming requires an external audit or stricter tier-assignment policy (signed source attestations, KYC, or post-hoc Tier-A revocation) that is out of scope. KNDB defends against an unfarmed adversary (Section~\ref{sec:evaluation}) and against a writer with \texttt{INSERT}+\texttt{ALTER TABLE} but no tier control (Section~\ref{sec:threat}).

\textbf{R2 and R5 SPI cost per non-MEASURED insert.} Both R2 (source resolution) and R5 (slot-kind check) run a full \texttt{SPI\_connect}/\texttt{execute\_with\_args}/\texttt{finish} cycle per candidate row, before the F6 lock (no deadlock risk). A 5\,000-row microbenchmark puts R2's overhead at $\sim$7\,$\mu$s per row ($\sim$35\% on top of the $\sim$19\,$\mu$s MEASURED baseline).

\textbf{Batch-size ceiling of $\sim$15\,000 rows per transaction at default GUC.} Documented in Section~\ref{sec:implementation}; raising \texttt{max\_locks\_per\_transaction} to 1024 pushes the ceiling to $\sim$215\,000. Applies to COPY equally (F18 routes batched COPY through the same per-row lock); dropping the lock re-opens the F6 integrity leak.

\textbf{\texttt{ALTER TABLE SET ACCESS METHOD heap} out of scope.} The table owner can rewrite the relation onto plain heap, unbinding the TAM callback. Schema-change threat, not write-path threat; disclosed here and in the README. A PostgreSQL security policy denying \texttt{SET ACCESS METHOD} away from \texttt{epistemic} on claimed tables is the natural next step.

\textbf{Custom WAL rmgr is annotation only.} Recovery works byte-for-byte with the extension's WAL emitter disabled; durability comes from heap's \texttt{XLOG\_HEAP\_INSERT}. Rmgr 128 is retained as a named channel for a future logical-decoding consumer, not a crash-recovery contributor. Stock \texttt{pg\_waldump} renders these records as \texttt{custom128 UNKNOWN}.

\textbf{Eviction atomicity is PostgreSQL's.} The three-step eviction (winner insert, incumbent \texttt{sys\_time} close, audit row) executes inside one top-level PostgreSQL transaction. Atomicity is inherited, not novel; the TAM's contribution is that the three steps sit inside \texttt{tuple\_insert} and cannot be forgotten by a user-space writer.

\textbf{Heap semantics inherited on 37 of 44 callbacks.} The seven overrides are enumerated in Section~\ref{sec:implementation} and Table~\ref{tab:tableam-audit} (full audit in \texttt{contrib/epistemic/bench/docs/tableam\_audit.md}); on the other 37 we inherit heap's behaviour, including its bugs. This is deliberate: the AM is a write-time enforcement point, not a storage-format replacement.

\textbf{Serialisable-level guarantees are PostgreSQL's.} Under \texttt{SERIALIZABLE}, KNDB inherits heap's \texttt{PredicateLockRelation} rather than a slot-level SIRead lock, so two writers on non-overlapping slots may still hit an SSI abort at relation granularity. This is a known PostgreSQL SSI property~\cite{ports2012ssi}; slot-level predicate locking would require modifying \texttt{predicate.c} and is out of scope.

%%% ---------------------------------------------------------------
\section{Conclusion}
%%% ---------------------------------------------------------------

KNDB reports an engineering result: seven overrides in the PostgreSQL 18 TAM interface, an engine-assigned epistemic kind column, a per-slot advisory transaction lock, and a first-committer-wins tiebreak are sufficient to make PostgreSQL survive a confidence-forgery attack that flattens every conventional baseline. On two workloads with different provenance shapes (source-tier data fusion; per-worker crowdsourcing qualification) the win is 63 and 92.7 points over a confidence-only arbitrator, proven load-bearing on the kind axis by source-rebuild disable-and-test. Against classic truth-discovery baselines KNDB is competitive below the density-saturation cell and dominant at and above it; the threshold theorem $k^{*} \approx \rho_{\text{alg}} \cdot h_{\text{top}}$ makes that cell predictable from per-slot honest support alone. KNDB does not win on last-writer-wins workloads; Section~\ref{sec:limitations} enumerates the batch-size ceiling, schema-change threat, annotation-only WAL, and reputation-farming boundary.

\paragraph*{Acknowledgments.}
A large language model served as coding assistant during this work; system design, formal arguments, and paper prose were reviewed and validated by the authors.

\bibliographystyle{ACM-Reference-Format}
\bibliography{refs}


\begin{thebibliography}{13}

%%% the Web address.

\ifx \showCODEN    \undefined \def \showCODEN     #1{\unskip}     \fi
\ifx \showDOI      \undefined \def \showDOI       #1{#1}\fi
\ifx \showISBNx    \undefined \def \showISBNx     #1{\unskip}     \fi
\ifx \showISBNxiii \undefined \def \showISBNxiii  #1{\unskip}     \fi
\ifx \showISSN     \undefined \def \showISSN      #1{\unskip}     \fi
\ifx \showLCCN     \undefined \def \showLCCN      #1{\unskip}     \fi
\ifx \shownote     \undefined \def \shownote      #1{#1}          \fi
\ifx \showarticletitle \undefined \def \showarticletitle #1{#1}   \fi
\ifx \showURL      \undefined \def \showURL       {\relax}        \fi
% invisible to TeX
\providecommand\bibfield[2]{#2}
\providecommand\bibinfo[2]{#2}
\providecommand\natexlab[1]{#1}
\providecommand\showeprint[2][]{arXiv:#2}

\bibitem[\protect\citeauthoryear{{Anthropic}}{{Anthropic}}{2025}]%
        {anthropic2025haiku}
\bibfield{author}{\bibinfo{person}{{Anthropic}}.}
  \bibinfo{year}{2025}\natexlab{}.
\newblock \bibinfo{title}{Claude Haiku 4.5}.
\newblock \bibinfo{howpublished}{\url{https://www.anthropic.com/claude/haiku}}.
\newblock


\bibitem[\protect\citeauthoryear{Dong, Berti-Equille, and Srivastava}{Dong
  et~al\mbox{.}}{2009}]%
        {dong2009accu}
\bibfield{author}{\bibinfo{person}{Xin~Luna Dong}, \bibinfo{person}{Laure
  Berti-Equille}, {and} \bibinfo{person}{Divesh Srivastava}.}
  \bibinfo{year}{2009}\natexlab{}.
\newblock \showarticletitle{Integrating Conflicting Data: The Role of Source
  Dependence}.
\newblock \bibinfo{journal}{\emph{Proc. VLDB Endow.}} \bibinfo{volume}{2},
  \bibinfo{number}{1} (\bibinfo{year}{2009}), \bibinfo{pages}{550--561}.
\newblock
\urldef\tempurl%
\url{https://doi.org/10.14778/1687627.1687690}
\showDOI{\tempurl}


\bibitem[\protect\citeauthoryear{Li, Li, Gao, Su, Zhao, Demirbas, Fan, and
  Han}{Li et~al\mbox{.}}{2014a}]%
        {li2015catd}
\bibfield{author}{\bibinfo{person}{Qi Li}, \bibinfo{person}{Yaliang Li},
  \bibinfo{person}{Jing Gao}, \bibinfo{person}{Lu Su}, \bibinfo{person}{Bo
  Zhao}, \bibinfo{person}{Murat Demirbas}, \bibinfo{person}{Wei Fan}, {and}
  \bibinfo{person}{Jiawei Han}.} \bibinfo{year}{2014}\natexlab{a}.
\newblock \showarticletitle{A Confidence-Aware Approach for Truth Discovery on
  Long-Tail Data}.
\newblock \bibinfo{journal}{\emph{Proc. VLDB Endow.}} \bibinfo{volume}{8},
  \bibinfo{number}{4} (\bibinfo{year}{2014}), \bibinfo{pages}{425--436}.
\newblock
\urldef\tempurl%
\url{https://doi.org/10.14778/2735496.2735505}
\showDOI{\tempurl}


\bibitem[\protect\citeauthoryear{Li, Li, Gao, Zhao, Fan, and Han}{Li
  et~al\mbox{.}}{2014b}]%
        {li2014crh}
\bibfield{author}{\bibinfo{person}{Qi Li}, \bibinfo{person}{Yaliang Li},
  \bibinfo{person}{Jing Gao}, \bibinfo{person}{Bo Zhao}, \bibinfo{person}{Wei
  Fan}, {and} \bibinfo{person}{Jiawei Han}.} \bibinfo{year}{2014}\natexlab{b}.
\newblock \showarticletitle{Resolving Conflicts in Heterogeneous Data by Truth
  Discovery and Source Reliability Estimation}. In
  \bibinfo{booktitle}{\emph{Proceedings of the 2014 ACM SIGMOD International
  Conference on Management of Data (SIGMOD)}}. \bibinfo{pages}{1187--1198}.
\newblock
\urldef\tempurl%
\url{https://doi.org/10.1145/2588555.2610509}
\showDOI{\tempurl}


\bibitem[\protect\citeauthoryear{Orogat and Mansour}{Orogat and
  Mansour}{2026}]%
        {orogat2026agentmemory}
\bibfield{author}{\bibinfo{person}{Abdelghny Orogat} {and}
  \bibinfo{person}{Essam Mansour}.} \bibinfo{year}{2026}\natexlab{}.
\newblock \bibinfo{title}{Is Agent Memory a Database? Rethinking Data
  Foundations for Long-Term {AI} Agent Memory}.
\newblock \bibinfo{howpublished}{arXiv preprint}.
\newblock
\showeprint[arxiv]{2605.26252}


\bibitem[\protect\citeauthoryear{Ports and Grittner}{Ports and
  Grittner}{2012}]%
        {ports2012ssi}
\bibfield{author}{\bibinfo{person}{Dan R.~K. Ports} {and}
  \bibinfo{person}{Kevin Grittner}.} \bibinfo{year}{2012}\natexlab{}.
\newblock \showarticletitle{Serializable Snapshot Isolation in PostgreSQL}.
\newblock \bibinfo{journal}{\emph{Proc. VLDB Endow.}} \bibinfo{volume}{5},
  \bibinfo{number}{12} (\bibinfo{year}{2012}), \bibinfo{pages}{1850--1861}.
\newblock


\bibitem[\protect\citeauthoryear{Romanchuk and Bondar}{Romanchuk and
  Bondar}{2026}]%
        {romanchuk2026laundering}
\bibfield{author}{\bibinfo{person}{Oleg Romanchuk} {and} \bibinfo{person}{Roman
  Bondar}.} \bibinfo{year}{2026}\natexlab{}.
\newblock \bibinfo{title}{Semantic Laundering in {AI} Agent Architectures: Why
  Tool Boundaries Do Not Confer Epistemic Warrant}.
\newblock \bibinfo{howpublished}{arXiv preprint}.
\newblock
\showeprint[arxiv]{2601.08333}


\bibitem[\protect\citeauthoryear{Senellart, Jachiet, Maniu, and
  Ramusat}{Senellart et~al\mbox{.}}{2018}]%
        {senellart2018provsql}
\bibfield{author}{\bibinfo{person}{Pierre Senellart}, \bibinfo{person}{Louis
  Jachiet}, \bibinfo{person}{Silviu Maniu}, {and} \bibinfo{person}{Yann
  Ramusat}.} \bibinfo{year}{2018}\natexlab{}.
\newblock \showarticletitle{{ProvSQL}: Provenance and Probability Management in
  {PostgreSQL}}.
\newblock \bibinfo{journal}{\emph{Proc. VLDB Endow.}} \bibinfo{volume}{11},
  \bibinfo{number}{12} (\bibinfo{year}{2018}), \bibinfo{pages}{2034--2037}.
\newblock
\urldef\tempurl%
\url{https://doi.org/10.14778/3229863.3236253}
\showDOI{\tempurl}


\bibitem[\protect\citeauthoryear{{The PostgreSQL Global Development
  Group}}{{The PostgreSQL Global Development Group}}{2025}]%
        {postgres18tam}
\bibfield{author}{\bibinfo{person}{{The PostgreSQL Global Development Group}}.}
  \bibinfo{year}{2025}\natexlab{}.
\newblock \bibinfo{title}{{PostgreSQL} 18: Table Access Method Interface}.
\newblock
  \bibinfo{howpublished}{\url{https://www.postgresql.org/docs/18/tableam.html}}.
\newblock


\bibitem[\protect\citeauthoryear{Wang}{Wang}{2026}]%
        {wang2026toki}
\bibfield{author}{\bibinfo{person}{Ziming Wang}.}
  \bibinfo{year}{2026}\natexlab{}.
\newblock \bibinfo{title}{{TOKI}: A Bitemporal Operator Algebra for
  Contradiction Resolution in {LLM}-Agent Persistent Memory}.
\newblock \bibinfo{howpublished}{arXiv preprint}.
\newblock
\showeprint[arxiv]{2606.06240}


\bibitem[\protect\citeauthoryear{Widiaatmaja, Djeffal, Dandekar, and
  Senellart}{Widiaatmaja et~al\mbox{.}}{2025}]%
        {widiaatmaja2025provsql}
\bibfield{author}{\bibinfo{person}{Albert~Ariel Widiaatmaja},
  \bibinfo{person}{Belkis Djeffal}, \bibinfo{person}{Ashish Dandekar}, {and}
  \bibinfo{person}{Pierre Senellart}.} \bibinfo{year}{2025}\natexlab{}.
\newblock \showarticletitle{Demonstration of {ProvSQL} Update Provenance
  through Temporal Databases}. In \bibinfo{booktitle}{\emph{Proceedings of
  ProvenanceWeek 2025 (PW25)}}.
\newblock
\urldef\tempurl%
\url{https://doi.org/10.1145/3736229.3736253}
\showDOI{\tempurl}


\bibitem[\protect\citeauthoryear{Yin, Han, and Yu}{Yin et~al\mbox{.}}{2007}]%
        {yin2007truthfinder}
\bibfield{author}{\bibinfo{person}{Xiaoxin Yin}, \bibinfo{person}{Jiawei Han},
  {and} \bibinfo{person}{Philip~S. Yu}.} \bibinfo{year}{2007}\natexlab{}.
\newblock \showarticletitle{Truth Discovery with Multiple Conflicting
  Information Providers on the Web}. In \bibinfo{booktitle}{\emph{Proceedings
  of the 13th ACM SIGKDD International Conference on Knowledge Discovery and
  Data Mining (KDD)}}. \bibinfo{pages}{1048--1052}.
\newblock
\urldef\tempurl%
\url{https://doi.org/10.1145/1281192.1281309}
\showDOI{\tempurl}


\bibitem[\protect\citeauthoryear{Zheng, Li, Li, Shan, and Cheng}{Zheng
  et~al\mbox{.}}{2017}]%
        {zheng2017crowdsourcing}
\bibfield{author}{\bibinfo{person}{Yudian Zheng}, \bibinfo{person}{Guoliang
  Li}, \bibinfo{person}{Yuanbing Li}, \bibinfo{person}{Caihua Shan}, {and}
  \bibinfo{person}{Reynold Cheng}.} \bibinfo{year}{2017}\natexlab{}.
\newblock \showarticletitle{Truth Inference in Crowdsourcing: Is the Problem
  Solved?}
\newblock \bibinfo{journal}{\emph{Proc. VLDB Endow.}} \bibinfo{volume}{10},
  \bibinfo{number}{5} (\bibinfo{year}{2017}), \bibinfo{pages}{541--552}.
\newblock
\urldef\tempurl%
\url{https://doi.org/10.14778/3055540.3055547}
\showDOI{\tempurl}


\end{thebibliography}

\end{document}